\documentclass[a4paper,fleqn]{cas-dc}

\usepackage[authoryear,longnamesfirst]{natbib}
\usepackage{cuted}
\usepackage{ upgreek }
\usepackage{amsthm}
\usepackage{url}
\usepackage{hyperref}

\DeclareMathOperator*{\argmin}{argmin}
\def\tsc#1{\csdef{#1}{\textsc{\lowercase{#1}}\xspace}}
\tsc{WGM}
\tsc{QE}
\tsc{EP}
\tsc{PMS}
\tsc{BEC}
\tsc{DE}
\newcommand{\tobs}{t_{\text{obs}}}
\newcommand{\tpred}{t_{\text{pred}}}

\newdefinition{rmk}{Remark}

\begin{document}
\let\WriteBookmarks\relax
\def\floatpagepagefraction{1}
\def\textpagefraction{.001}
\shorttitle{Predicting Network Dynamics Using Sparse Identification and Proper Orthogonal Decomposition}

\shortauthors{Luo et~al.}

\title [mode = title]{Unraveling Low-Dimensional Network Dynamics: A Fusion of Sparse Identification and Proper Orthogonal Decomposition}                      
\tnotemark[1]

\tnotetext[1]{
The  work  described  in  this  paper  was  substantially supported  by  a  grant  from 
City University of Hong Kong (Project No. 9610639).}

\author[1]{Rui Luo}[type=editor,
                        orcid=0000-0003-0711-8039]

\cormark[1]

\ead{ruiluo@cityu.edu.hk}

\credit{Conceptualization of this study, Methodology, Software}

\affiliation[1]{organization={City University of Hong Kong},
    addressline={83 Tat Chee Ave, Kowloon Tong}, 
    city={Hong Kong SAR}
    }

\cortext[cor1]{Corresponding author}

\begin{abstract}
This study addresses the challenge of predicting network dynamics, such as forecasting disease spread in social networks or estimating species populations in predator-prey networks. Accurate predictions in large networks are difficult due to the increasing number of network dynamics parameters that grow with the size of the network population (e.g., each individual having its own contact and recovery rates in an epidemic process), and because the network topology is unknown or cannot be observed accurately. 

Inspired by the low-dimensionality inherent in network dynamics, we propose a two-step method. First, we decompose the network dynamics into a composite of principal components, each weighted by time-dependent coefficients. Subsequently, we learn the governing differential equations for these time-dependent coefficients using sparse regression over a function library capable of describing the dynamics.
We illustrate the effectiveness of our proposed approach using simulated network dynamics datasets. The results provide compelling evidence of our method's potential to enhance predictions in complex networks.

\end{abstract}


\maketitle

\section{Introduction}
The study of dynamics in networks has gained significant attention due to its potential applications in various domains.  These include the spread of infectious diseases through contact networks \citep{anderson1992may, pastor2015epidemic}, the formation of the glass ceiling effects and segregation in social networks \citep{luo2023mutual, luo2022controlling}, the transmission of information in communication networks \citep{hill2010emotions}, and the propagation of misinformation in social networks \citep{luo2022mitigating}, and synchronization processes within power grids \citep{rohden2012self}.

However, predicting network dynamics is a challenging task due to the following factors: 

\noindent (1) {\bf Network structure-dynamics interplay:}
The combination of network structures and dynamical processes can give rise to nonlinear phenomena \citep{skardal2020higher, d2019explosive}. In addition, the interaction term in the coupled ordinary differential equations (ODEs) governing network dynamics results in a high-dimensional parameter space, which complicates parameter estimation \citep{wu2019parameter}. 

\noindent (2) {\bf Size and complexity of real-world networks:} Many real-world networks comprise a substantial  number of nodes. This results in an increase in the number of model parameters, leading to computational challenges \citep{mikkelsen2017learning} and potentially stiff systems \citep{schweiger2020modeling}. For instance, in the context of epidemic processes on networks, parameters like curing rates and infection probabilities augment linearly with the node count. Consequently, predicting dynamics in large networks can be computationally demanding and time-consuming.

\noindent (3) {\bf Uncertainty in the network structure:} The actual structure of a network, which represents the connections between its nodes, is often either unknown or difficult to observe with precision. For example, in multi-agent dynamical systems like dynamic traffic network, node features evolve over time, and the relationships among agents can also vary, where new edges form and existing edges drop.  Such uncertainty in the network structure can significantly impact the accuracy of network dynamics prediction.
Existing techniques for network dynamics prediction often either disregard the network structure or assume a fixed topology \citep{huang2021coupled}. This neglects the dynamic nature of contact patterns and potential changes in network structure, which are crucial for accurate predictions. For example, the assumption of a fixed topology may not hold in scenarios where individuals change their contact patterns due to social distancing measures or behavior changes.

\begin{figure*}
	\centering
	\includegraphics[width=1\textwidth]{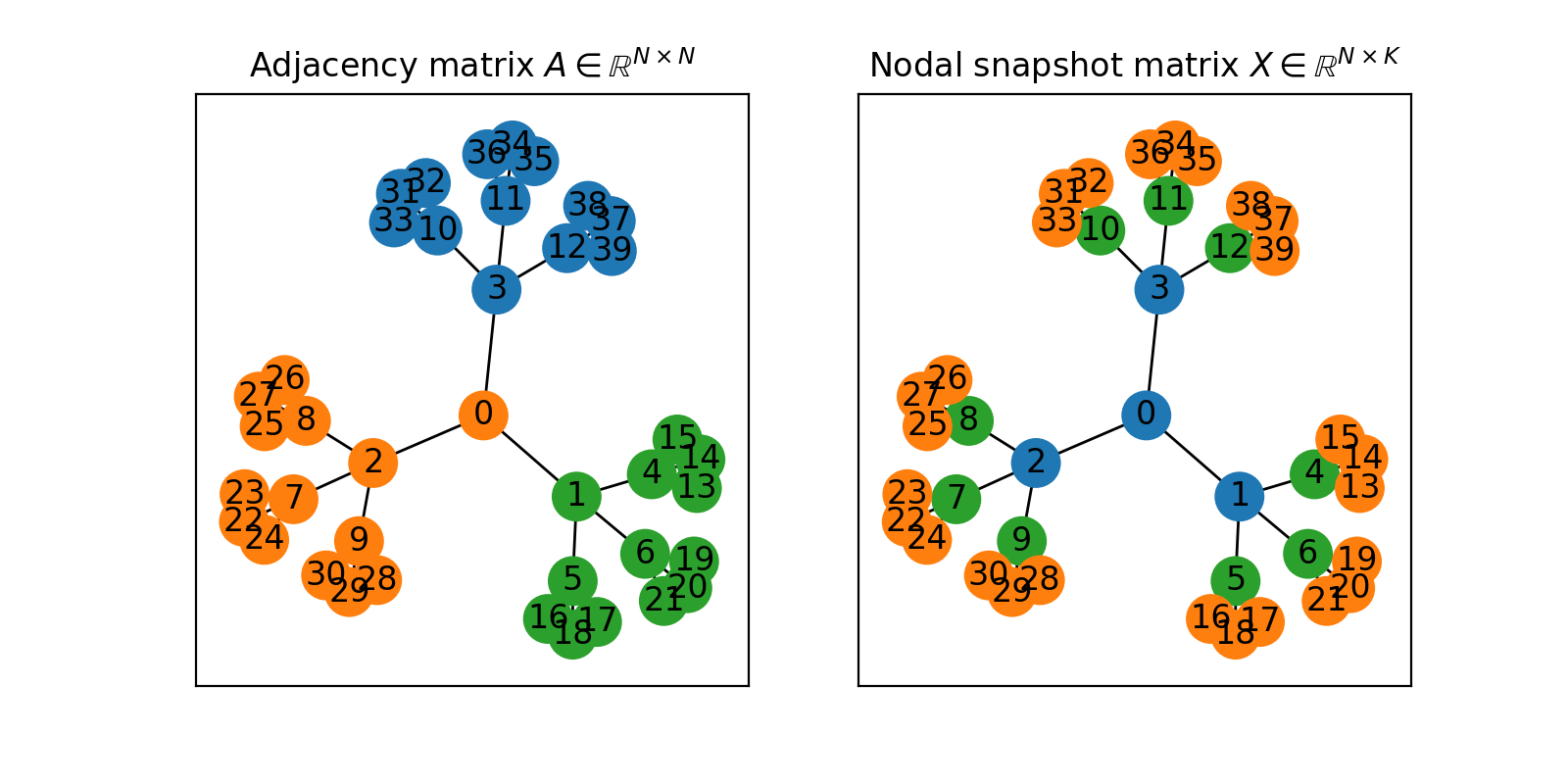}
	\caption{This figure illustrates the results of applying spectral clustering to a tree graph undergoing a Susceptible-Infected-Susceptible (SIS) process (\ref{subsec: SIS}). Different colors represent distinct groups partitioned by spectral clustering. In the left subfigure, spectral clustering is applied to the adjacency matrix $A$ (\ref{eq: adj}), uncovering the community structure of the network. Conversely, in the right subfigure, spectral clustering is applied to the network snapshot matrix $X$ (\ref{eq: SVD}), which records the network dynamics observed over a time interval. This approach identifies different stages of epidemic propagation, offering insights that could enhance future predictions.}
	\label{fig: spectral clustering}
\end{figure*}

Despite the challenges entailed in predicting network dynamics, as previously mentioned, it is worth investigating the structure inherent in these dynamics. This structure can prove instrumental in guiding predictions, even in the absence of knowledge about the network topology. An illustrative example is presented in Figure \ref{fig: spectral clustering}. Here, spectral clustering \citep{von2007tutorial} is applied to nodes within a tree network which is undergoing an Susceptible-Infected-Susceptible (SIS) epidemic process (refer to Section \ref{subsec: SIS} for details). The spectral clustering is implemented using either the adjacency matrix of the corresponding graph or the node snapshot matrix, as defined in Section \ref{subsec: POD}.

The results reveal that using the adjacency matrix leads to the partitioning of nodes into distinct communities, which uncovers the community structure within the network. On the other hand, using the node snapshot matrix allows for the identification of nodes at varying stages of epidemic propagation (even without the knowledge of network structure), which uncovers the patterns inherent in the network dynamics. This observation serves as a motivation for using Proper Orthogonal Decomposition (POD) on the observed node dynamics to facilitate future predictions.

In this study, we propose an approach to address the challenges in predicting network dynamics by leveraging the low-dimensionality inherent in these dynamics \citep{prasse2022predicting}. We use the POD method to express network dynamics as a combination of agitation modes, weighted by a vector of time-dependent coefficients. To enhance the predictive capability of our approach, we apply Sparse Identification of Nonlinear Dynamics (SINDy) to the resulting time-dependent coefficients. This allows us to uncover the low-rank structures inherent in the network dynamics.

Our proposed approach has several advantages over existing techniques. Firstly, it captures the dynamic nature of contact patterns and changes in the network structure, leading to more accurate predictions. Secondly, it reduces the dimensionality of the prediction task, making it computationally more efficient. Finally, it provides a more robust prediction by leveraging the average nodal states in each cluster. This approach has the potential to significantly improve predictions of epidemic dynamics in networks and can be applied to various domains beyond epidemiology.

\section{Network Dynamics Formulation}
This section introduces the formulation of network dynamics. We then consider a concrete example of modelling of the epidemic dynamics on a network by considering a Susceptible-Infected-Susceptible (SIS) process on a network. 

\subsection{Modelling dynamics on networks as a coupled ODE}
Network dynamics, which aims at describing the progression and interplay among interconnected agents, can be effectively captured and articulated using the framework of coupled ordinary differential equations (ODEs) \citep{prasse2022predicting}. 

Consider a network modeled as a graph consisting of $N$ nodes and an $N \times N$ adjacency matrix $A$. Each element $a_{ij}$ in the matrix is defined as follows:
\begin{equation}\label{eq: adj}
a_{ij}
\begin{cases}
> 0, & \text{if there is a directed link from node $j$ to node $i$;} \\
= 0, & \text{otherwise}. 
\end{cases}
\end{equation}

The evolution of node states is described by the following coupled ODE: 
\begin{equation}\label{eq: ode general}
    \frac{dx_i(t)}{dt} = f_i(x_i(t)) + \sum_{j=1}^{N} a_{ij} g(x_i(t), x_j(t)),
\end{equation}
where $x_i(t)$ represents node $i$'s state at $t$, $f_i(x_i(t))$ describes the self-dynamics and $g(x_i(t), x_j(t))$ represents the interaction. We can represent the state vector of all nodes as  $x(t) = (x_1(t), \cdots, x_N(t))^T \in \mathbb{R}^{N}$. Thus, we can express equation (\ref{eq: ode general}) more compactly:
\begin{equation}\label{eq: ode matrix}
    \frac{dx(t)}{dt} = f_A( x(t)),
\end{equation}
where the subscript ${}_A$ of the coupling function $f_A$ indicates the graph coupled interactions in the ODE system.  

In the context of recent advancements in graph machine learning, $f_{A}$ can be selected as a graph neural network module, such as the graph convolution operator \citep{kipf2016semi}:
\begin{equation}
f_{A}(x(t)) = \hat{D}^{-\frac{1}{2}}\hat{A}\hat{D}^{-\frac{1}{2}}x(t) W,
\end{equation}
where $\hat{A}=A+I$ signifies the adjacency matrix augmented with self-loops, $\hat{D}$ corresponds to the associated diagonal degree matrix, and $W$ is a learnable weight matrix.

A generalization of (\ref{eq: ode matrix}) to second-order network dynamics, termed the Graph-Coupled Oscillator Network (GraphCON) \citep{rusch2022graph}, uses nonlinear oscillators interconnected by a network to represent network dynamics:
\begin{equation}
\frac{d^2x_i(t)}{dt^2} = \sigma\Big(\sum_{j \in \mathcal{N}_i} f_{A}\big(x_i(t), x_j(t)\big) \Big) - \gamma x_i(t) - \alpha \frac{dx_i(t)}{dt},
\end{equation}
in which $\mathcal{N}_i$ denotes the neighborhood of node $i$. Written as a matrix-vector product, this equation becomes:
\begin{equation}
\frac{d^2x(t)}{dt^2} = \sigma \big( f_{A}(x(t)) \big) - \gamma x(t) - \alpha \frac{d x(t)}{dt}.
\end{equation}

In the subsequent discussion, we primarily concentrate on first-order network dynamics, as detailed in equations (\ref{eq: ode general}, \ref{eq: ode matrix}). We delve into a specific instance of an epidemic process on networks in Section \ref{subsec: SIS}.

\subsection{Susceptible-infected-susceptible epidemics on a network}\label{subsec: SIS}
Epidemic models typically categorize populations based on disease stages \citep{anderson1992may}. The SIS model uses a two-state system, labeling individuals as susceptible (S) or infected (I). It permits two transitions: $S\rightarrow I$, when a susceptible person becomes infected, and $I\rightarrow S$, when an infected person recovers and returns to susceptibility. The SIS model assumes no immunity, so individuals can repeatedly cycle through the $S\rightarrow I\rightarrow S$ process.

In recent years, an impressive research effort has been devoted to understanding the effects of complex network topologies on the SIS model \citep{pastor2015epidemic}. The focus on complex network topologies is essential due to the inherent intricacies in real-world networks, which can impact the dynamics of disease spread.

In the representation of (\ref{eq: ode general}), $a_{ij}$ denotes the infection rate from node $j$ to node $i$. Each node in the network can be either infected or susceptible. The nodal state $x_i(t) \in [0, 1]$ represents the probability of node $i$ being infected at time $t$. Additionally, every node is associated with a constant curing rate $\delta_i > 0$.

The dynamics of $x_i(t)$ can thus be described as:
\begin{equation}\label{eq: SIS}
\frac{dx_i(t)}{dt} = - \delta_i x_i(t) + \sum_{j=1}^{N} a_{ij} (1 - x_i(t)) x_j(t)
\end{equation}
This equation describes how the probability of node $i$ being infected changes over time, taking into account the curing rate $\delta_i$ and the influence of adjacent nodes in the network. Writing (\ref{eq: SIS}) in a matrix-vector product form:
\begin{equation}
    \frac{dx(t)}{dt} = - \delta \odot x(t) + A(1-x(t)) \odot x(t),
\end{equation}
where $\delta = (\delta_1, \cdots, \delta_N)^T$ denotes the curing rate vector of all nodes, and $\odot$ represents elementwise multiplication.

In this study, our primary objective is to forecast the values of $x(t)$ for times $t > \tobs$, where $\tobs$ denotes the observation time. In other words, we aim to predict the network dynamics by leveraging the data gathered over a specific period. In Section \ref{sec: prediction}, we demonstrate how to achieve this objective by exploiting the low-dimensionality of the network dynamics.

\section{Low-Dimensional Network Dynamics}\label{sec: prediction}
We demonstrate that the interplay of the dynamics and network structure can be approximated using a low-dimensional representation via the Proper Orthogonal Decomposition (POD).  Then, using Sparse Identification of Nonlinear Dynamics (SINDy), we demonstrate how the time-dependent coefficients of the POD modes can be predicted without a precise knowledge of the network structure. 

\subsection{Dimensionality reduction of network dynamics using POD}\label{subsec: POD}
The Proper Orthogonal Decomposition (POD) method was originally formulated in the field of turbulence research with the aim of identifying coherent structures within chaotic fluid flows \citep{lumley1967pod}. By analyzing a few dominant POD modes, referred to as agitation modes, researchers aimed to uncover organizing flow patterns that are otherwise obscured in the complexity of turbulence.

To predict network dynamics, it is often simpler to (1) \textbf{reduce its dimensionality} and (2) \textbf{predict the emerging low-dimensional dynamics}. For the initial step, we draw inspiration from \citet{prasse2022predicting}, using POD to reduce the dimension of network dynamics. 

Specifically, at any time $t$, POD provides an approximation of the $N \times 1$ nodal state vector $x(t)$ using only $m \ll N$ dominant modes:
\begin{equation}\label{eq: POD}
x(t) \approx \sum_{p=1}^{m} c_p(t) y_p
\end{equation}
where the agitation modes $y_1, \ldots, y_m$ are the orthonormal vectors that do not vary with time. The time-dependent coefficients $c_p(t)$ are computed by projecting $x(t)$ onto each agitation mode:
\begin{equation} \label{eq: cp}
c_p(t) = y_p^T x(t), p=1, \cdots, m.
\end{equation}
By reducing the dynamics to the dominant $m$ modes, POD dramatically simplifies the prediction task compared to modeling the full $N$-dimensional state vector.

To obtain the agitation modes $y_p$, we use $K$ equidistant nodal state observations with the observation time interval $[0, \tobs]$. These observations are concatenated to form a nodal state matrix, or snapshot matrix \citep{brunton2019data}, $X = (x(0), x(\Delta t), \ldots, x((K-1) \Delta t) )$. The agitation modes are then derived as the first $m$ left-singular vectors of $X$:
\begin{equation}\label{eq: SVD}
X = U \Sigma V^T.
\end{equation}
Here, $U$ represents an $N \times N$ orthogonal matrix, $\Sigma$ is an $N \times K$ rectangular diagonal matrix, and $V$ is an $K \times K$ orthogonal matrix. The non-zero diagonal elements of $\Sigma$ are arranged in descending order, i.e., $\sigma_1 \geq \ldots \geq \sigma_K \geq 0$. The agitation modes $y_1 = U_{\cdot 1}, \dots, y_m = U_{\cdot m}$ correspond to the first $m$ columns of $U$.

\citet{prasse2022predicting} illustrate that the POD modes extracted from network snapshots observed within a time interval $[0, \tobs]$ remain precise at future times $t > \tobs$. Consequently, Consequently, the challenge of predicting network dynamics is essentially reduced to forecasting the time-dependent coefficients $c_p(t)$, as denoted in (\ref{eq: cp}). We present our solution in the following Section \ref{subsec: SINDy}.

\begin{rmk}[ODE-reducible renewal equation]
As a brief diversion, we will now discuss the connection between POD and ODE-reducible renewal equations \citep{diekmann2018finite}. A renewal equation is ODE-reducible if its solution can be fully recovered from the solution of a system of linear ODEs. %

Consider a finite dimensional state representation of Markovian population (compartment) models in epidemic process. Structured populations with $n$ individual states can be modelled as a system of ODEs:
\begin{equation} \label{eq: population ode}
    \frac{dN(t)}{dt} = (B+H)N(t),
\end{equation}
where the $i$-th component of $N(t)$ represents the density of individuals within state $i$. The $n\times n$ matrix $H$ generates the process of state transition: $H_{ij}, i\neq j$ is the rate at which an individual jumps from state $i$ to state $j$, and $-H_{jj}$ is the rate at which an individual leaves state $j$; and the $n\times n$ matrix $B$ represents reproduction: $B_{ij}$ is the rate at which an individual in state $j$ gives birth to an individual in state $i$.

It is noted that in many population models, the possible states at birth form a proper subset of all states. That is, the dimension $k$ of the range of $B$ is typically less than $n$. This observation bears a resemblance to the low-dimensional structure inherent in network dynamics. Accordingly, $B$ can be decomposed as $B=VU^T$ where $V \in \mathbb{R}^{n \times k}$, and (\ref{eq: population ode}) becomes:
\begin{equation} \label{eq: population ode 2}
    \frac{dN(t)}{dt} = (VU^T + H) N(t).
\end{equation}
Define the birth rate vector $b(t) \in \mathbb{R}^k$ at time $t$ by:
\begin{equation} \label{eq: population birth rate}
    b(t) = U^T N(t),
\end{equation}
(\ref{eq: population ode 2}) can then be written as:
\begin{equation}
    \frac{dN(t)}{dt} = Vb(t) + HN(t).
\end{equation}
Applying the variation of constants formula yields:
\begin{equation} \label{eq: population variation of constants}
    N(t) = \int_{-\infty}^{t} e^{(t-s)H} Vb(s) ds
\end{equation}
Substituting (\ref{eq: population variation of constants}) into (\ref{eq: population birth rate}) leads to the renewal equation that $b$ satisfies:
\begin{equation}
    b(t) = \int_{-\infty}^{t} U^T e^{(t-s)H} Vb(s) ds.
\end{equation}
$b(t)$ is equivalent to the time-dependent coefficients $c(t)$ in the POD approximation (\ref{eq: POD}). 
\end{rmk}

\subsection{Sparse identification of dynamics of time-dependent coefficients}\label{subsec: SINDy}
To discover the evolution pattern of the time-dependent coefficients $c_p(t), p=1, \cdots, m$, we use the Sparse Identification of Nonlinear Dynamics (SINDy), which uncovers the underlying nonlinear dynamics from the available data  \citep{brunton2016discovering, oishi2023nonlinear}. 
The key idea underneath SINDy is that a dynamic system
\begin{equation}
    \frac{dx(t)}{dt}=f(x)
\end{equation}
often exhibits sparsity within a specific function space. 

SINDy
characterizes the differential equation governing the time-dependent coefficients, or POD mode amplitudes as illustrated in \citet{oishi2023nonlinear}, as a sparse combination drawn from a library of potential functions that could depict the dynamics. Consider the ODE for the evolution of $c(t) = (c_1(t), \cdots, c_m(t))^T$:
\begin{equation}\label{eq: dc(t)}
    \frac{dc(t)}{dt} = h(c(t)), 
\end{equation}

To determine the function $h$ from data, we collect $c(t)$ for the time interval $0 \leq t \leq \tobs$. And we approximate the derivative $\frac{dc(t)}{dt}$ numerically by differencing $c(t)$. 

Similar to the snapshot matrix introduced in Section \ref{subsec: POD}, we organize the time history of $c(t)$ into a matrix:
\begin{equation}
\begin{split}
    C & = \begin{bmatrix}
        c(0) & c(\Delta t) & \cdots & c((K-1)\Delta t)
    \end{bmatrix} \\
    & = \begin{bmatrix}
        c_1(0) & c_1(\Delta t) & \cdots & c_1((K-1)\Delta t) \\
        c_2(0) & c_2(\Delta t) & \cdots & c_2((K-1)\Delta t) \\
        \vdots & \vdots & \ddots & \vdots \\
        c_m(0) & c_m(\Delta t) & \cdots & c_m((K-1)\Delta t)
    \end{bmatrix}\in \mathbb{R}^{m\times K}.
\end{split}
\end{equation}
And the corresponding history of time derivatives is:
\begin{equation}
\begin{split}
    \dot{C} & = \begin{bmatrix}
        \dot{c}(0) & \dot{c}(\Delta t) & \cdots & \dot{c}((K-1)\Delta t)
    \end{bmatrix} \\
    & = \begin{bmatrix}
        \dot{c}_1(0) & \dot{c}_1(\Delta t) & \cdots & \dot{c}_1((K-1)\Delta t) \\
        \dot{c}_2(0) & \dot{c}_2(\Delta t) & \cdots & \dot{c}_2((K-1)\Delta t) \\
        \vdots & \vdots & \ddots & \vdots \\
        \dot{c}_m(0) & \dot{c}_m(\Delta t) & \cdots & \dot{c}_m((K-1)\Delta t)
    \end{bmatrix} \in \mathbb{R}^{m\times K},
\end{split}
\end{equation}
where $\dot{c}_i(k\Delta t) \approx \frac{c_i((k+1)\Delta t) - c_i((k-1)\Delta t)}{2 \Delta t}$ is computed using central difference approximation. 

Next, a library of candidate functions $\Theta(C)$ is constructed as follows:
\begin{equation}
    \Theta(C) = 
    \begin{bmatrix}
        \vert & \vert & \vert & & \vert & \vert & \\
        1 & C & C^{P_2} & \cdots & \sin{C} & \cos{C} & \cdots \\
        \vert & \vert & \vert & & \vert & \vert & 
    \end{bmatrix}^T \in \mathbb{R}^{d\times K},
\end{equation}
where $ C^{P_2}$ denotes second-order polynomials:

\begin{strip}
\begin{equation}
    C^{P_2} = 
    \begin{bmatrix}
        c_1^2(0) & c_1(0)c_2(0) & \cdots & c_2^2(0) & \cdots & c_m^2(0) \\
        c_1^2(\Delta t) & c_1(\Delta t)c_2(\Delta t) & \cdots & c_2^2(\Delta t) & \cdots & c_m^2(\Delta t) \\
        \vdots & \vdots & \ddots & \vdots & \ddots & \vdots \\
        c_1^2((K-1)\Delta t) & c_1((K-1)\Delta t)c_2((K-1)\Delta t) & \cdots & c_2^2((K-1)\Delta t) & \cdots & c_m^2((K-1)\Delta t)
    \end{bmatrix} \in \mathbb{R}^{K \times d_{P_2}}.
\end{equation}
\end{strip}
Here, $d_{P_2} = m^2$ denotes the number of second-order polynomials used as candidate functions, and $d$ denotes the total number of candidate functions. Typically there are more data samples than functions \citep{brunton2016discovering}, i.e., $d < K$.

Each row of $\Theta(C)$ signifies a candidate function, while each column of $C$ represents the dynamics of a specific time-dependent coefficient. SINDy assumes that only a few of the candidate functions will influence the dynamics. To this end, SINDy formulates a sparse regression problem to determine the sparse matrix of coefficients $\Upxi \in \mathbb{R}^{m\times d}$, which will isolate the terms from the library that impact the dynamics:
\begin{equation} \label{eq: sparse regression}
    \argmin\limits_{\Upxi} \| \dot{C} - \Upxi \Theta(C)  \|_2 + \gamma \|\Upxi\|_1,
\end{equation}
where $\|\Upxi\|_1$ is $l_1$ norm used for regularization, and $\gamma$ is a regularization parameter. Each row $\xi_p$ of $\Upxi$ is a sparse vector with each of its elements signifying whether the corresponding candidate function is active for the dynamics of $c_p(t)$. 
(\ref{eq: sparse regression}) is then solved using the Sparse Relaxed Regularized Regression (SR3) algorithm \citep{zheng2018unified}.

The objective of using SINDy is to identify the governing equation of the dynamics of the time-dependent coefficients $c_p(t), p=1, \cdots, m$. In Remark \ref{rmk: consensus}, we demonstrate through an example of network consensus dynamics that the governing equation of the time-dependent coefficients admits a simple form and allows for effective prediction.

\begin{rmk}[A consensus dynamic example]\label{rmk: consensus}
Take the network consensus dynamics \citep{saber2003acc} as an example, where $f_i(x_i(t))=0, g(x_i(t), x_j(t))=x_j(t) - x_i(t)$, then (\ref{eq: ode general}) becomes:
\begin{equation}
    \frac{dx_i(t)}{dt} = \sum_{j=1}^{N} a_{ij} (x_j(t) - x_i(t)),
\end{equation}
and in matrix-vector product form
\begin{equation}
    \frac{dx(t)}{dt} = -L x(t),
\end{equation}
where $L = D - A$ is the Laplacian matrix. 
According to \citet{lyapunov1992general}, the ODE above is asymptotically stable when all eigenvalues of $L$ are negative. And the trajectories of all $x_i(t)$ will be close enough as $t$ grows sufficiently large. 

For an undirected graph, $L$ is diagonalizable with real eigenvalues: $L = U \Lambda U^T$, where $U=(u_1, \cdots, u_N)$ is an orthogonal matrix, and $\Lambda = \text{diag}(\lambda_1, \cdots, \lambda_N)$ is a diagonal matrix with entries $0 \leq \lambda_1 \leq \cdots \leq \lambda_N$. Substitute $x(t) = U y(t) $,
\begin{equation}
    \frac{dy(t)}{dt} = -U^T L U y(t) = - \Lambda y(t),
\end{equation}
thus
\begin{equation} \label{eq: solution to Laplacian dynamics}
    x(t) = Uy(t) = (u_1, \cdots, u_N) 
    \begin{bmatrix}
        w_1 e^{-\lambda_1 t} \\
        \vdots \\
        w_N e^{-\lambda_N t}
    \end{bmatrix} = \sum_{i=1}^{N} u_i w_i e^{-\lambda_i t},
\end{equation}
and 
\begin{equation}
    x_i(t) = \sum_{j=1}^{N} u_j^{(i)} w_j e^{-\lambda_j t},
\end{equation}
where $u_j$ denotes the $i$-th entry of the eigenvector $u_j$, and $w_j$ is the constant which satisfies the initial condition $\sum_{j=1}^{N} u_j^{(i)} w_j = x(0)$. In the summation of $N$ exponentials, as $t$ increases, the exponential with higher order will dominate. We can thus approximate (\ref{eq: solution to Laplacian dynamics}) as:
\begin{equation}\label{eq: consensus dynamics}
    x(t) = \sum_{i=1}^{m} u_i w_i e^{-\lambda_i t},
\end{equation}
where $m < N$ denotes the number of exponential functions with the smallest negative exponents, which correspond to the $m$ smallest eigenvalues of $L$.

Comparing (\ref{eq: consensus dynamics}) with (\ref{eq: POD}), the time-dependent coefficients can be approximated as:
\begin{equation}
    c_p(t) = e^{-\lambda_p t},   
\end{equation}
which leads to
\begin{equation}
    \dot{C} = 
    \begin{bmatrix}
    -\lambda_{1} & & \\
    & \ddots & \\
    & & -\lambda_{m}
    \end{bmatrix}
    C.
\end{equation}
Then, using the terminology in SINDy, we can effectively represent the library of candidate functions and the sparse coefficient matrix as
\begin{equation}
    \Theta(C) = C, \quad \Upxi = \text{diag}(-\lambda_1, -\lambda_2, \cdots, -\lambda_m).
\end{equation}
\end{rmk}

This simple example illustrates that the sparse output derived from SINDy accurately corresponds with the underlying nonlinear dynamics of the time-dependent coefficients.

\subsection{Network dynamics prediction using SINDy} \label{subsec: prediction}
With the sparse coefficient matrix $\Upxi$ obtained via the SINDy approach in Section \ref{subsec: SINDy}, we illustrate the prediction of network dynamics in future time periods.

The differential equation of time-dependent coefficients (\ref{eq: dc(t)}) resulted from SINDy (\ref{eq: sparse regression}) is:
\begin{equation}
    \frac{dc(t)}{dt} = h(c(t)) = \Upxi \Theta(c(t)),
\end{equation}
where $\Theta(c(t)) \in \mathbb{R}^{d}$ is a vector as opposed to $\Theta(C)$, which is a data matrix. 

Subsequently, the time-dependent coefficients at a future time instance, denoted as $\tpred$, can be computed through integration as follows:
\begin{equation}
c(\tpred) = c(\tobs) + \int_{\tobs}^{\tpred} \frac{dc(t)}{dt} dt
\end{equation}

Lastly, the predicted nodal state at time $\tpred$ can be expressed as:
\begin{equation} \label{eq: prediction}
x(\tpred) = \sum_{p=1}^{m} c_p(\tpred) y_p,
\end{equation}
where the time-invariant agitation modes $y_p$ are obtained by singular value decomposition of the snapshot matrix $X$ (\ref{eq: SVD}).

\begin{rmk}[Prediction via a surrogate network]\label{rmk: surrogate}

We present an alternative approach to predicting network dynamics that uses a surrogate adjacency matrix, denoted as $\hat{A}$. While this approach produces accurate results when applied to observational data, it is contingent upon having knowledge of the parameters governing the dynamics. 

As per Proposition 2 in \citet{prasse2022predicting}, the system of linear equations (\ref{eq: POD})—which approximates the nodal state vector $x(t)$ using the first $m$ agitation modes—results in an underdetermined system. This implies there exists an infinite number of adjacency matrices that can accurately predict the network dynamics.

To resolve this issue, one can opt for L1-regularized least squares optimization to compute the surrogate adjacency matrix, denoted as $\hat{A}$. The derivative in equation (\ref{eq: SIS}) can be approximated using a difference equation:
\begin{equation}
\begin{split}
\frac{x_i((k+1)\Delta t) - x_i(k \Delta t)}{\Delta t} & \approx -\delta_i x_i(k\Delta t) \\
& + \sum_{j=1}^{N} a_{ij} (1 - x_i(k\Delta t) ) x_j(k \Delta t)
\end{split}
\end{equation}

Afterwards, the surrogate adjacency matrix $\hat{A}$ can be determined by solving the corresponding optimization problem:
\begin{equation}
\begin{split}
&\argmin_{\hat{a}_{i1}, \ldots, \hat{a}_{iN}} \sum_{k=0}^{K-1} \Big(\frac{x_i((k+1)\Delta t) - x_i(k \Delta t)}{\Delta t} + \delta_i x_i(k \Delta t) \\
& \phantom{====} - \sum_{j=1}^{N} a_{ij} (1-x_i(k \Delta t))x_j(k \Delta t) \Big)^2 + \rho_i \sum_{j=1}^{N} \hat{a}_{ij}, \\
&\text{s.t.} \quad \hat{a}_{ij} \geq 0, j=1, \ldots, N,
\end{split}
\end{equation}
for each node $i$. 
Then, the future nodal state vector can be estimated using $\hat{A}$:
\begin{equation}
    x(\tpred) = x(\tobs) +  \int_{\tobs}^{\tpred} -\delta \odot x(t) + \hat{A}(1-x(t)) \odot x(t) dt.
\end{equation}

However, the practical application of this method is constrained by its dependence on accurate knowledge of the dynamics, such as the curing rate $\delta_i$ of nodes.
\end{rmk}

\begin{figure*}
	\centering
	\includegraphics[width=1\textwidth]{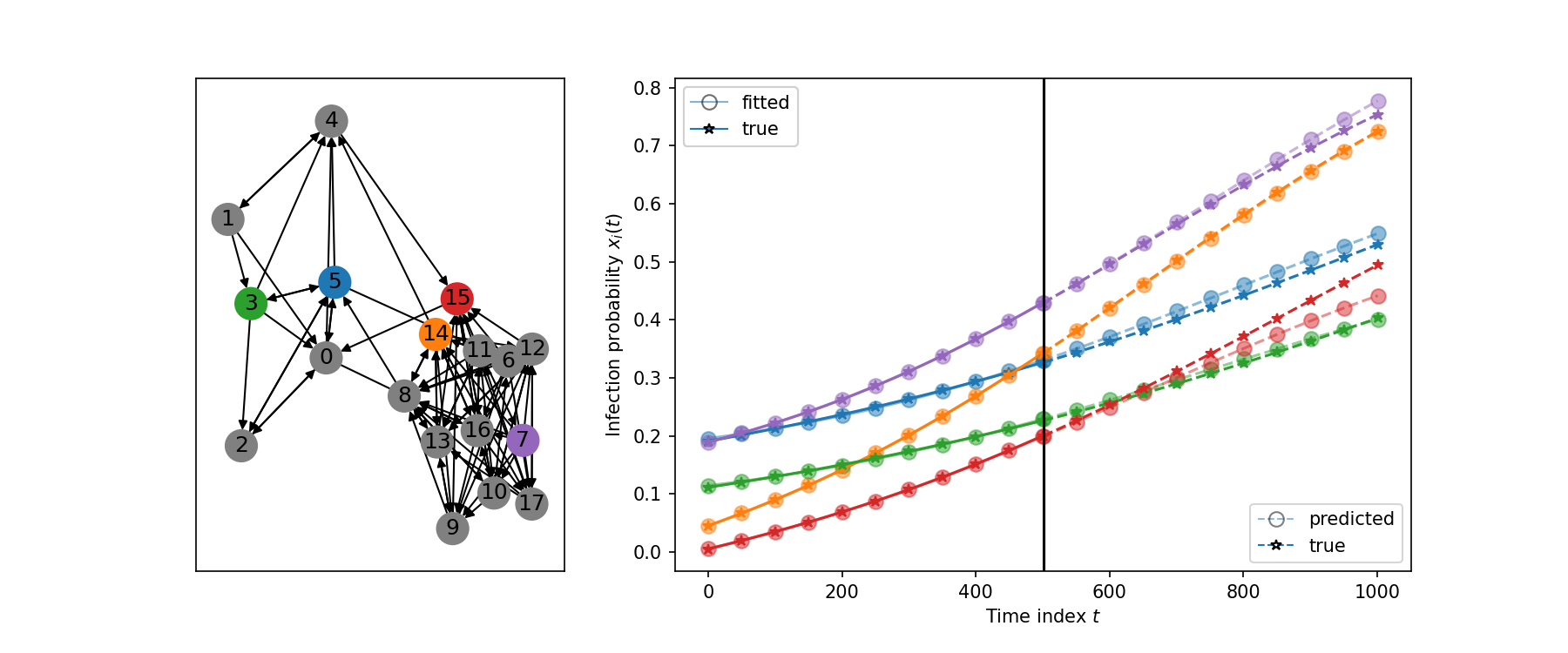}
	\caption{We present the prediction results of a SIS process (Section \ref{subsec: SIS}) on a directed SBM using our proposed method (Section \ref{subsec: prediction}). The first half of the discretized time-series data is used to form the network snapshot matrix. The fitted network dynamics in thefirst half are computed using 2 POD agitation modes, expressed as $x_{\text{fit}}(t) = \sum_{p=1}^{2} c_p(t) y_p$, where $0 \leq t \leq \tobs$. The predicted network dynamics for a random selection of 5 nodes are shown in the second half.}
	\label{fig: sbm}
\end{figure*}

\section{Numerical Study}
To validate the efficacy of our proposed method (Section \ref{subsec: prediction}) in predicting epidemic dynamics in networks (Section \ref{subsec: SIS}), we conduct experiments on simulated and real-world social networks. 

For simulating the epidemic process, we randomly initialize the infection probability of each node $x_i(0)$ uniformly between 0 and 0.2. Similarly, we assign each node with a curing rate $\delta_i$ uniformly at random between 0 and 0.2. 

We first apply our approach to a directed network sampled from the stochastic block model (SBM), which reflects the homophily and community structure often present in real-world networks. Subsequently, we test our approach on Zachary’s Karate Club graph \citep{zachary1977information} and Florentine families graph \citep{breiger1986cumulated}. 

We use the first half of the data as the observation to construct the network snapshot matrix. This matrix is then utilized to formulate the agitation modes $y_p$ and the time-dependent coefficients $c_p(t)$ (\ref{eq: cp}) via POD (Section \ref{subsec: POD}).
The fitted data, given by $\hat{x}(t) = \sum_{p=1}^{m} c_p(t) y_p, t \in [0, \tobs]$, is plotted alongside the true observational data.

Subsequently, we predict the differential equation (\ref{eq: dc(t)}) that governs the dynamics of the time-dependent coefficients $c_p(t)$, using SINDy (Section \ref{subsec: SINDy}). The resulting prediction is computed using (\ref{eq: prediction}).

\begin{figure*}
	\centering
	\includegraphics[width=1\textwidth]{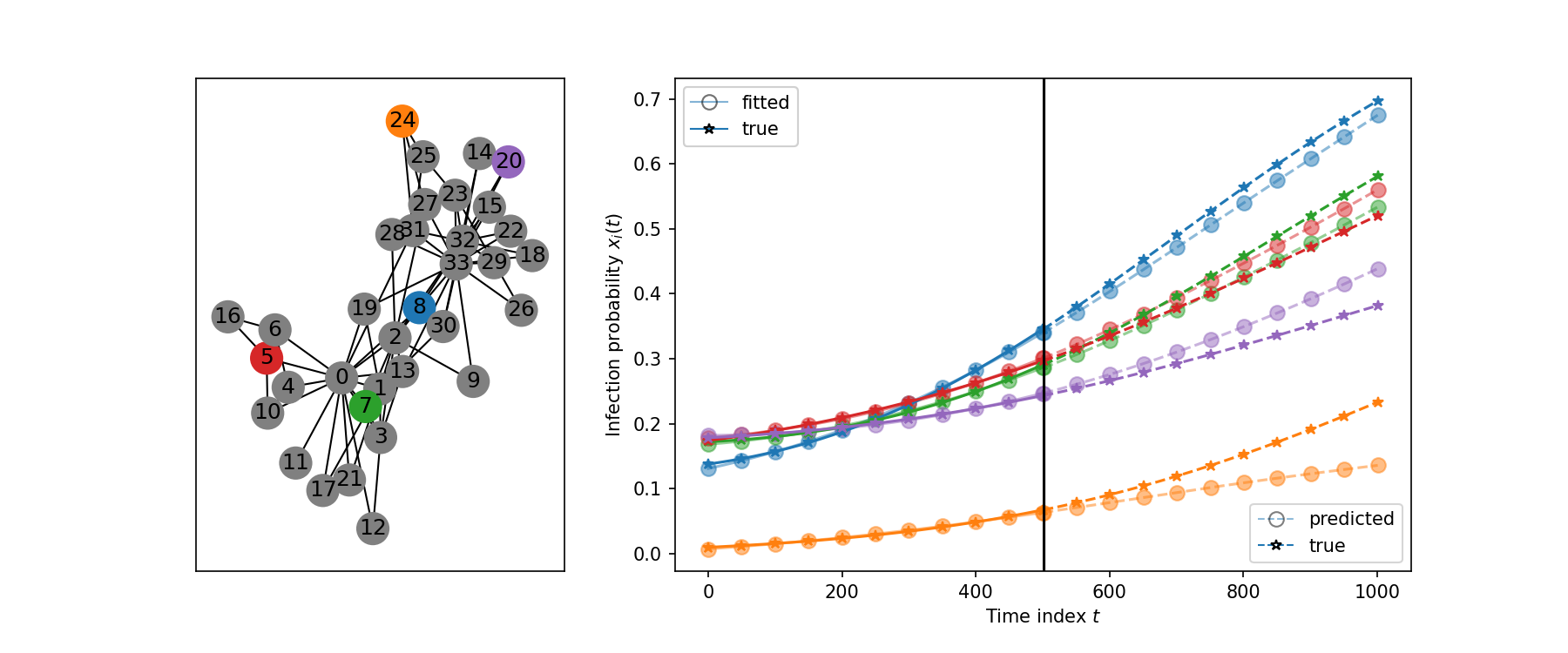}
	\caption{Prediction results of a SIS process on Zachary’s Karate Club graph.}
	\label{fig: karate}
\end{figure*}

\begin{figure*}
	\centering
	\includegraphics[width=1\textwidth]{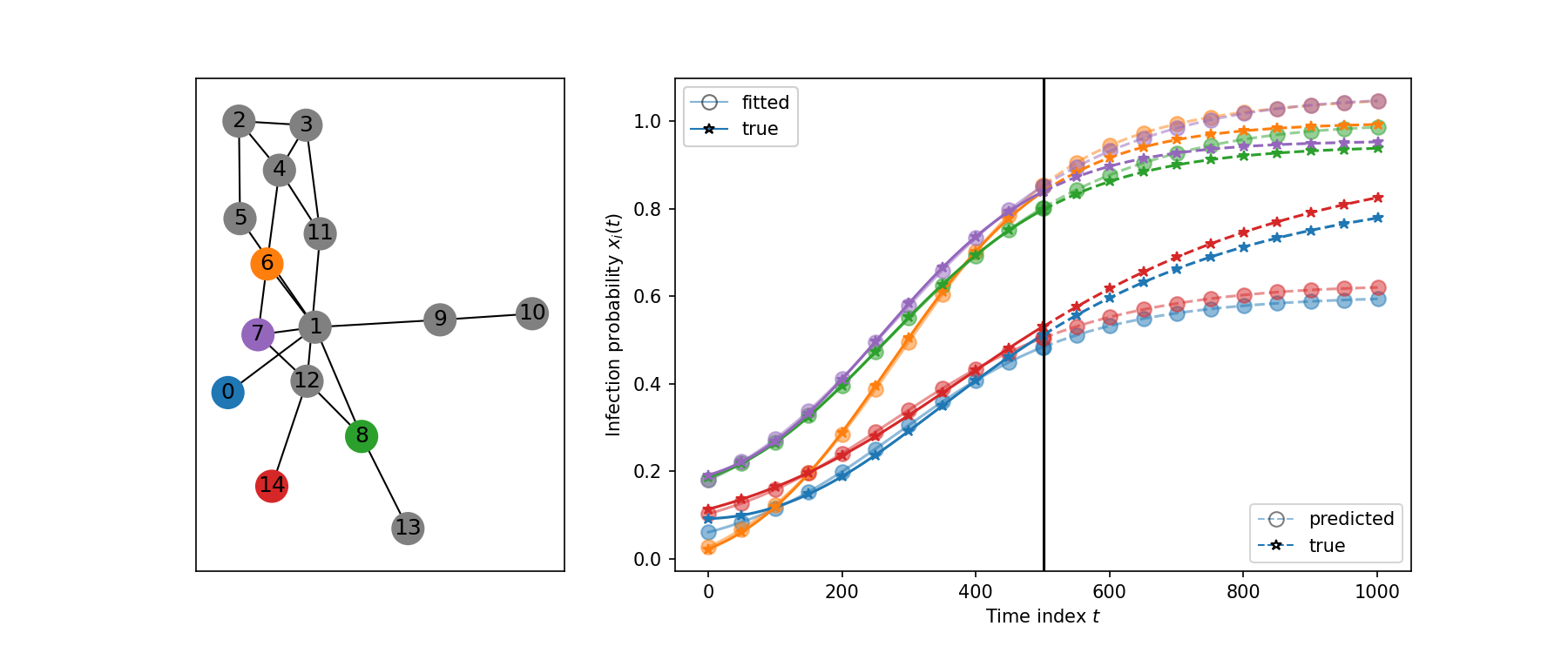}
	\caption{Prediction results of a SIS process on Florentine families graph.}
	\label{fig: florentine}
\end{figure*}

Our results in Fig. \ref{fig: sbm}, \ref{fig: karate}, and \ref{fig: florentine} show that the proposed method achieves satisfactory prediction results. This supports our hypothesis that exploiting the low-dimensional structure inherent in the dynamics of the time-dependent coefficients can enhance the accuracy of epidemic dynamics predictions in networks.

In summary, our experiments substantiate the efficacy of our proposed approach in predicting epidemic dynamics within networks, even without knowledge of the network structure or the parameters of the dynamics. Future studies could extend this approach to networks with time-varying topologies and assess the influence of different network topologies and model parameters on prediction accuracy.

\section{Related Works} 

Machine learning models specifically designed for graph structures have been developed for learning and predicting the network dynamics. 
\citet{huang2021coupled} uses graph neural networks (GNNs) to model the interactions between nodes and predict the evolution of node features. To accommodate the dynamic nature of network structure, they propose the use of a graph neural network to represent the coupled Ordinary Differential Equations (ODEs) pertaining to both latent node embeddings and network edges. 
This is represented by the following equations:
\begin{equation}
\begin{split}
    \frac{dz_{i\rightarrow j}(t)}{dt} &= f_e(z_i(t) \| z_j(t) \| z_{i\rightarrow j}(t)), \\
    A_{ij}(t) &= f_{edge2value}(z_{i\rightarrow j}(t)), \\
    \frac{dZ(t)}{dt} &= \sigma(A(t)Z(t)W) - Z(t) + Z(0), 
\end{split}
\end{equation}
where $z_i(t)$ denotes node $i$'s latent embedding, $z_{i\rightarrow j}(t)$ denotes the directed interaction term, or influence, from node $i$ to node $j$, and $\sigma(\cdot)$ is a graph convolution layer followed by a nonlinear activation function. 

\citet{zhuang2020ordinary} proposed Graph ODE (GODE) which generalizes Laplacian smoothing to a continous smoothing process:
\begin{equation}
    \frac{dx(t)}{dt} = -\gamma \hat{D}^{-\frac{1}{2}} \hat{L} \hat{D}^{-\frac{1}{2}} x(t),
\end{equation}
where $\gamma$ is a positive scaling constant, and the nonnegativity of the eigenvalues of the symmetrically normalized Laplacian ensures the stability of the dynamics. 

Graph-Coupled Oscillator Network (GraphCON) \citep{rusch2022graph} uses ODEs to model network dynamics with dampened, controlled non-linear oscillators linked by a network. Consider a second-order network dynamics: %
\begin{equation}
    \frac{d^2x_i(t)}{dt^2}  = \sigma\Big(\sum_{j \in \mathcal{N}_i} f_{\theta, A}\big(x_i(t), x_j(t)\big) \Big) - \gamma x_i(t) - \alpha \frac{dx_i(t)}{dt},
\end{equation}
where $\mathcal{N}_i$ denotes node $i$'s neighborhood, and $f_{\theta, A}$ is a coupling function with learnable parameter $\theta$ and the adjacency matrix $A$. Expressing the equation as a matrix-vector product,
\begin{equation}
    \frac{d^2x(t)}{dt^2} = \sigma \big( f_{\theta, A}(x(t)) \big) - \gamma x(t) - \alpha \frac{d x(t)}{dt}.
\end{equation}
The coupling function $f_{\theta, A}$ can be chosen as any graph neural network module, such as the graph convolution operator \citep{kipf2016semi}:
\begin{equation}
    f_{\theta, A}(x(t)) = \hat{D}^{-\frac{1}{2}}\hat{A}\hat{D}^{-\frac{1}{2}}x(t) W,
\end{equation}
where $\hat{A}=A+I$ denotes the adjacency matrix with self-loops, and $\hat{D}$ denotes the corresponding diagonal degree matrix. 

\citet{perraudin2017stationary} used the graph localization operator to define the wide sense stationarity of graph signals. They proved that stationary graph signals are characterized by a well-defined power spectral density that can be estimated following a Wiener-type estimation procedure. They noticed that stationary graph signal possesses the nice property that Laplacian eigenvectors are similar to the covariance eigenvectors.

The primary constraint of the existing approaches is their reliance on the accurate knowledge of the network's adjacency matrix. This dependence critically limits their applicability in situations where the network structure is either inaccurately observed or susceptive to changes; for instance, scenarios involving the edge deletion or rewiring. 

Another line of research relevant to this study is the network reconstruction problem based on the observed dynamics. Essentially, this problem aims to discover the underlying connection pattern based on a multivariate time series, or multivariate temporal point processes (MTPP). 
\citet{hallac2017network} considered the dynamic network inference as an optimization problem on a chain graph, where each node objective solves for a network slice at each timestamp using graphical Lasso, and edge objectives define the penalties that enforce temporal consistency. They also developed various penalty functions which encode diffrent chane behaviors of the network structure. Their approach differs from our method which also predict the values given previous data, that is, can explain as well as predict network behavior
\citet{belilovsky2017learning} addressed the undirected network inference problem by mapping empirical covariance matrices to estimated graph structures. 

\citet{wu2020connecting} proposed the following neural network model to learn an adjacency matrix from node features. They reduce the computational cost by adopting a sampling-based algorithm that randomly split the nodes into several groups and only learns a sugraph structure based on the sampled nodes. They also extracted uni-directional relationships among nodes, illustrated as the following equations:
\begin{equation}
\begin{split}
    M_1 & = \text{tanh}(E_1 \Theta_1), \\
    M_2 & = \text{tanh}(E_2 \Theta_2), \\
    A & = \text{ReLU}(\text{tanh}(M_1 M_2^T - M_2 M_1^T)),
\end{split}
\end{equation}
where $E_1$ and $E_2$ represent randomly initialized node embeddings, $\Theta_1$ and $\Theta_2$ are learnable weight matrices, and the subtraction and ReLU activation regularize the adjacency matrix so that if $A_{ij}$ is positive, its diagonal counterpart $A_{ji}$ will be zero. We extend their method to learn a directed (instead of uni-directional) adjacency matrix and network dynamics from a multivariate time series data (instead of static node features). 

\citet{khademi2020deep} introduces a graph neural network model that generates a scene graph for an image. The method starts with a complete probabilistic graph network and then uses a Q-learning framework to select nodes sequentially. The model then conducts pruning to refine the graph structure. This approach has shown promise in accurately capturing the relationships between objects in an image and generating a structured representation of the scene.
In their recent work, \citet{murphy2021deep} propose a novel graph neural network architecture that uses the attention mechanism to predict epidemic dynamics on a network. This approach is able to capture the complex relationships between nodes, leading to accurate predictions of epidemic spread. %

Machine learning algorithms have also been employed in the classification of network dynamics. One notable application was conducted by \citet{cheng2014can}, who utilized logistic regression to predict the growth trajectory of resharing cascades on social media networks. Through their research, they discovered that temporal and structural features outperformed content, original poster, and resharer features as significant indicators of whether a cascade would continue to grow.
\citet{bassi2022learning} focused on a binary classification problem, aiming to determine if a network system adhering to specific dynamics will either synchronize or converge to a non-synchronizing limit cycle. Their findings underscored the value of integrating graph statistics with several iterations of network dynamics, a combination that remarkably enhanced accuracy in their predictions.

\section{Conclusions}
Many current approaches for predicting network dynamics depend on knowledge of both the network structure and the parameters of the dynamics. This reliance creates unrealistic prerequisites for real-world predictions. To circumvent this issue, we've proposed an innovative approach that capitalizes on the low-dimensionality inherent in network dynamics.

Our approach begins by implementing a Proper Orthogonal Decomposition (POD) over the observed network snapshots. This step yields the agitation modes and the time-dependent coefficients. Subsequently, our method employs a sparse regression framework, the Sparse Identification of Nonlinear Dynamics (SINDy), to identify the dynamical process governing the evolution of the time-dependent coefficients.

Simulation results on both simulated and real-world networks demonstrate that our proposed algorithm yields satisfactory outcomes. This supports the feasibility and effectiveness of our approach for real-world application.

% \bibliographystyle{cas-model2-names}

% \bibliography{cas-refs}

\end{document}